\documentclass[aps,prd,floats,twocolumn,epsf,nofootinbib,showpacs]{revtex4}
\usepackage{latexsym}
\usepackage{amssymb}
\usepackage[dvips]{graphicx}
\usepackage[applemac]{inputenc}
\usepackage[nottoc]{tocbibind} 
\usepackage{float}
\usepackage{fancyhdr}
\usepackage{amsfonts}
\usepackage{amsmath}
\usepackage{color}
\usepackage{mathrsfs}
\usepackage{bm}
\usepackage{epsfig}

\newcommand{\be}{\begin{equation}}
\newcommand{\ee}{\end{equation}}

\def\lsim{\mathrel{\raise.3ex\hbox{$<$\kern-.75em\lower1ex\hbox{$\sim$}}}}

\def\gsim{\mathrel{\raise.3ex\hbox{$>$\kern-.75em\lower1ex\hbox{$\sim$}}}}

\def\beq{\begin{eqnarray}}
\def\eeq{\end{eqnarray}}
\def\bea{\begin{eqnarray}}
\def\eea{\end{eqnarray}}

\begin{document}

\title{Prospects For Identifying Dark Matter With CoGeNT}

\author{Chris Kelso$^{a}$ and Dan Hooper$^{b,c}$}

\address{
$^a$Department of Physics, University of Chicago, Chicago, IL 60637 \\
$^b$Center for Particle Astrophysics, Fermi National Accelerator
Laboratory, Batavia, IL 60510 \\ 
$^c$Department of Astronomy and Astrophysics, University of Chicago,
Chicago, IL 60637}

\begin{abstract}

It has previously been shown that the excess of events reported by the CoGeNT collaboration could be generated by elastically scattering dark matter particles with a mass of approximately 5-15 GeV. This mass range is very similar to that required to generate the annual modulation observed by DAMA/LIBRA and the gamma rays from the region surrounding the Galactic Center identified within the data of the Fermi Gamma Ray Space Telescope. To confidently conclude that CoGeNT's excess is the result of dark matter, however, further data will likely be needed. In this paper, we make projections for the first full year of CoGeNT data, and for its planned upgrade. Not only will this body of data more accurately constrain the spectrum of nuclear recoil events, and corresponding dark matter parameter space, but will also make it possible to identify seasonal variations in the rate. In particular, if the CoGeNT excess is the product of dark matter, then one year of CoGeNT data will likely reveal an annual modulation with a significance of 2-3$\sigma$. The planned CoGeNT upgrade will not only detect such an annual modulation with high significance, but will be capable of measuring the energy spectrum of the modulation amplitude. These measurements will be essential to irrefutably confirming a dark matter origin of these events.

\end{abstract}

\pacs{95.36.+x; FERMILAB-PUB-10-457-A}

\maketitle


\section{Introduction}

In early 2010, the CoGeNT collaboration reported that they had observed on the order of 100 events above expected backgrounds with ionization energies in the range of approximately 0.4 to 1.0 keV~\cite{cogent}. While the origin of these events is not yet certain, it has been shown that they could be accounted for with dark matter with a mass in the range of 5-15 GeV and an elastic scattering cross section with nuclei of $\sim$$10^{-4}$ pb ($\sim$$10^{-40}$ cm$^2$)~\cite{zurek,compare}.

The CoGeNT excess is particularly intriguing when compared to other signals potentially resulting from dark matter. For nearly a decade, the DAMA collaboration (and more recently, the DAMA/LIBRA collaboration) has reported an annual modulation of their event rate, and has interpreted this behavior as evidence for particle dark matter. According to the most recent DAMA/LIBRA results, which make use of
over 1.17 ton-years of data, they observe a modulation with a significance of 8.9$\sigma$, and with a phase consistent with that predicted for elastically scattering dark matter~\cite{damanew}. The simplest interpretation of this signal which does not conflict with the null results of other dark matter searches~\cite{cdms,xenon100} introduces light dark matter particles ($\lsim 10$ GeV) with an elastic scattering cross section with nuclei on the order of $\sim$$10^{-4}$ pb~\cite{petriello}. After all constraints~\cite{cdms,xenon100,cdmslow,constraints} and uncertainties~\cite{uncertain,uncertain2} are taken into account, a region of dark matter parameter space exists in which both the CoGeNT and DAMA/LIBRA signals can be accommodated~\cite{consistent}.

More recently, an anomalous component of gamma rays from the inner 0.5$^{\circ}$ around the Galactic Center was identified in the first two years of data from the Fermi Gamma Ray Space Telescope (FGST)~\cite{Hooper:2010mq}. The spectrum of this signal peaks at 2-4 GeV (in $E^2$ units) and is highly concentrated around the Galactic Center (but is not point-like). Although no known astrophysical sources or mechanism can account for this emission, annihilating dark matter can provide a good fit to the observed spectrum and morphology of the signal. If interpreted as dark matter annihilation products, the observed gamma rays imply a dark matter mass in the range of 7.3 to 9.2 GeV, very similar to that required to accommodate both CoGeNT and DAMA/LIBRA.

Although the evidence collectively provided by CoGeNT, DAMA/LIBRA, and FGST has become fairly compelling, more information will be needed before a confident and conclusive claim of discovery can be made. The annual modulation reported by DAMA/LIBRA, however, is based on data taken over 13 years, and little new information is expected from DAMA/LIBRA in the foreseeable future. Similarly, the gamma ray signal from the Galactic Center was extracted from two full years of FGST data (August 2008-August 2010). In contrast, the excess reported by CoGeNT was observed over a period of only 56 days in late 2009. With the currently existing data that has been collected over the past year, the CoGeNT collaboration will be able to dramatically improve upon their measurement on the nuclear recoil spectrum. Furthermore, if elastically scattering dark matter is responsible for their observed excess, CoGeNT will likely be sensitive to the predicted annual modulation. Looking further into the future, a planned expansion of CoGeNT (four 0.9 kg detectors, with an estimated combined fiducial mass of $\sim$2.5 kg) will not only be capable of detecting the presence of a modulation, but will be sensitive to the energy spectrum of the modulation, providing an irrefutable confirmation that their excess originates from dark matter.

The remainder of this paper is structured as follows. In Sec.~\ref{current}, we calculate the spectrum of events from elastically scattering dark matter in the CoGeNT detector and describe the current status of the CoGeNT excess. In Sec.~\ref{future}, we will project the ability of CoGeNT to constrain the mass and cross section of the dark matter using their first full year of data, and with the planned CoGeNT upgrade, focusing on the ability of CoGeNT to detect the annual modulation of the dark matter signal. We argue that this will provide an essential confirmation of the dark matter origin of the observed excess. In Sec.~\ref{others}, we discuss the role that other direct detection experiments will likely play in studying dark matter in the region implied by CoGeNT and DAMA/LIBRA. In Sec.~\ref{summary} we summarize our results and conclusions.

\section{CoGeNT and Elastically Scattering Dark Matter}
\label{current}

CoGeNT, located in the Soudan Underground Laboratory in Northern Minnesota, consists of a 475 gram target mass of Germanium (a fiducial mass of 330 grams). Although this is considerably smaller mass than is employed by CDMS, XENON, and other direct detection experiments, CoGeNT's very low backgrounds at low recoil energies make it exceptionally sensitive to low mass WIMPs.

The spectrum (in nuclear recoil energy) of dark matter induced elastic scattering events is given by~\cite{ls}
\be
\frac{dR}{dE_R} = N_T \frac{\rho_{DM}}{m_{DM}} \int_{|\vec{v}|>v_{\rm
min}} d^3v\, vf(\vec{v},\vec{v_e}) \, \frac{d\sigma}{d E_R},
\label{rate1}
\ee
where $N_T$ is the number of target nuclei, $m_{DM}$ is the mass of
the dark matter particle, $\rho_{DM}$ is the local dark matter
density, $\vec{v}$ is the dark matter velocity in the frame of the
Earth, $\vec{v_e}$ is the velocity of the Earth with respect 
to the galactic halo, and $f(\vec{v},\vec{v_e})$ is the distribution
function of dark matter particle velocities, which we take to be 
the standard Maxwell-Boltzmann distribution:
\be
f(\vec{v},\vec{v_e}) = \frac{1}{(\pi v_0^2)^{3/2}} {\rm
e}^{-(\vec{v}+\vec{v_e})^2/v_0^2}.
\ee
The Earth's speed relative to the galactic halo is given by
$v_e=v_{\odot}+v_{\rm orb}{\rm cos}\,\gamma\, {\rm
cos}[\omega(t-t_0)]$ where $v_{\odot}=v_0+12\,{\rm km/s}$, 
$v_{\rm orb}=30 {\rm km/s}$, ${\rm cos}\,\gamma=0.51$, $t_0={\rm June
\, 2nd}$, and $\omega=2\pi/{\rm year}$. We will consider values of $v_0$ over a range of 180 to 320 km/s and values of the galactic escape velocity between 460 and 640 km/s~\cite{velocity}. Note that the minimum dark matter velocity required to impart a recoil of energy, $E_R$, is given by $v_{\rm min} = \sqrt{E_R m_N/2 \mu^2}$, where $m_N$ is the mass of the target nucleus and $\mu$ is the reduced mass of the dark matter particle and the target nucleus. Throughout this study, we will take the local dark matter density to be $\rho_{DM}=0.3$ GeV/cm$^3$.

As the germanium isotopes which make up the CoGeNT detector contain little net spin, we are forced to consider spin-independent interactions to generate the observed events. In this case, we have
\be
\frac{d\sigma}{d E_R} = \frac{m_N}{2 v^2} \frac{\sigma_n}{\mu_n^2}
\frac{\left[f_p Z+f_n (A-Z)\right]^2}{f_n^2} F^2(q),
\label{cross1}
\ee
where $\mu_n$ is the reduced mass of the dark matter particle and
nucleon (proton or neutron), 
$\sigma_n$
 is the scattering cross section of the dark matter 
particle with 
neutrons,
$Z$ and $A$ are the atomic and mass numbers of the nucleus, and
$f_{n,p}$ are the coupling strengths of the dark matter particle to
neutrons and protons respectively. The nuclear form factor, $F(q)$, accounts for the finite momentum
transfer in scattering events.  In our calculations, we adopt the
Helm form factor:
\begin{equation}
F(q) = \frac{3 j_1(q R_1)}{q R_1} \,\, e^{-{\frac{1}{2}q^2 s^2}}, 
\end{equation}
where $j_1$ is the second spherical Bessel function and $R_1$ is
given by
\begin{equation}
R_1 = \sqrt{c^2 + \frac{7 \pi^2 a^2}{3} - 5 s^2}.
\end{equation}
Here, $c \approx 1.23 A^{1/3} -0.60$ fm, $a \approx 0.523$ fm, and $s
\approx 0.9$ fm have been determined by fits to nuclear physics
data~\cite{Gondolo,Fricke}. 

To convert from nuclear recoil energy to the measured ionization energy, we have to scale the results by the appropriate quenching factor. Over the energy range of the CoGeNT signal (approximately 0.4 to 2 keVee, where keVee denotes keV electron equivalent), the quenching factors for germanium have been well measured~\cite{Ge,Ge2} and are described by $Q_{\rm Ge}(E_{\rm Recoil}=3\,{\rm keV})=0.218$, and with the energy dependence predicted by the Lindhard theory (see Ref.~\cite{consistent}).

\begin{figure}[t]
\centering
{\includegraphics[angle=0.0,width=3.5in]{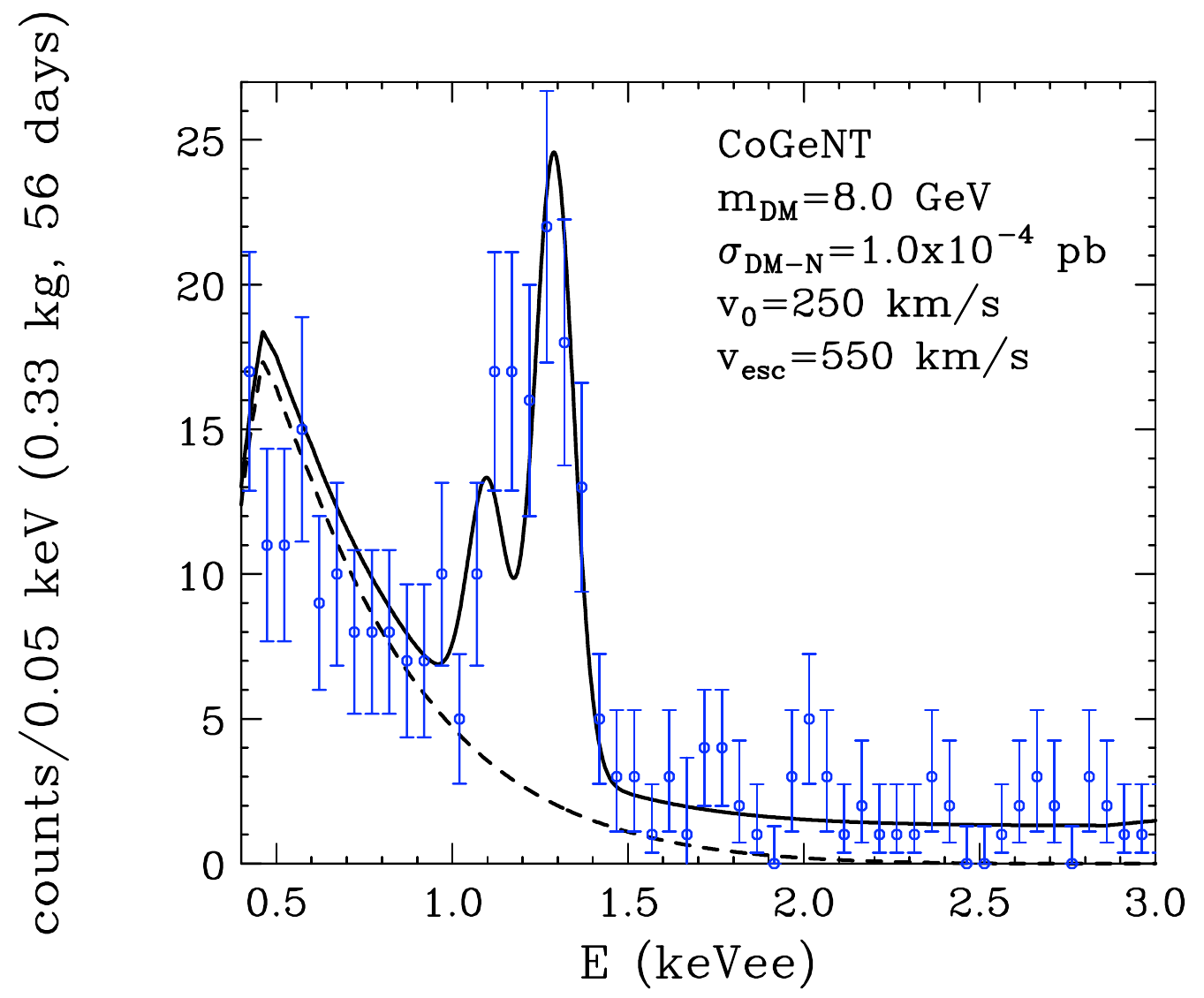}}
\caption{The current spectrum of events reported by CoGeNT compared to the spectrum predicted for an elastically scattering dark matter particle. The dashed line denotes the spectrum of dark matter events alone, while the solid line is the dark matter spectrum plus backgrounds.}
\label{currentspec}
\end{figure}

In Fig.~\ref{currentspec}, we show the current results from CoGeNT, compared to the spectrum predicted for an 8 GeV dark matter particle with an elastically scattering cross section with nucleons of $10^{-4}$ pb and a velocity distribution described by $v_0=250$ km/s and $v_{\rm esc}=550$ km/s. For the background spectrum, we consider a simple flat distribution of events, and Gaussian peaks at 1.1 and 1.29 keV with widths given by the resolution of the detector, which arise from well understood physics (see Ref.~\cite{cogent}). We multiply the overall spectrum by the efficiency factors as described in Ref.~\cite{cogent}.

\begin{figure}[t]
\centering
{\includegraphics[angle=0.0,width=3.5in]{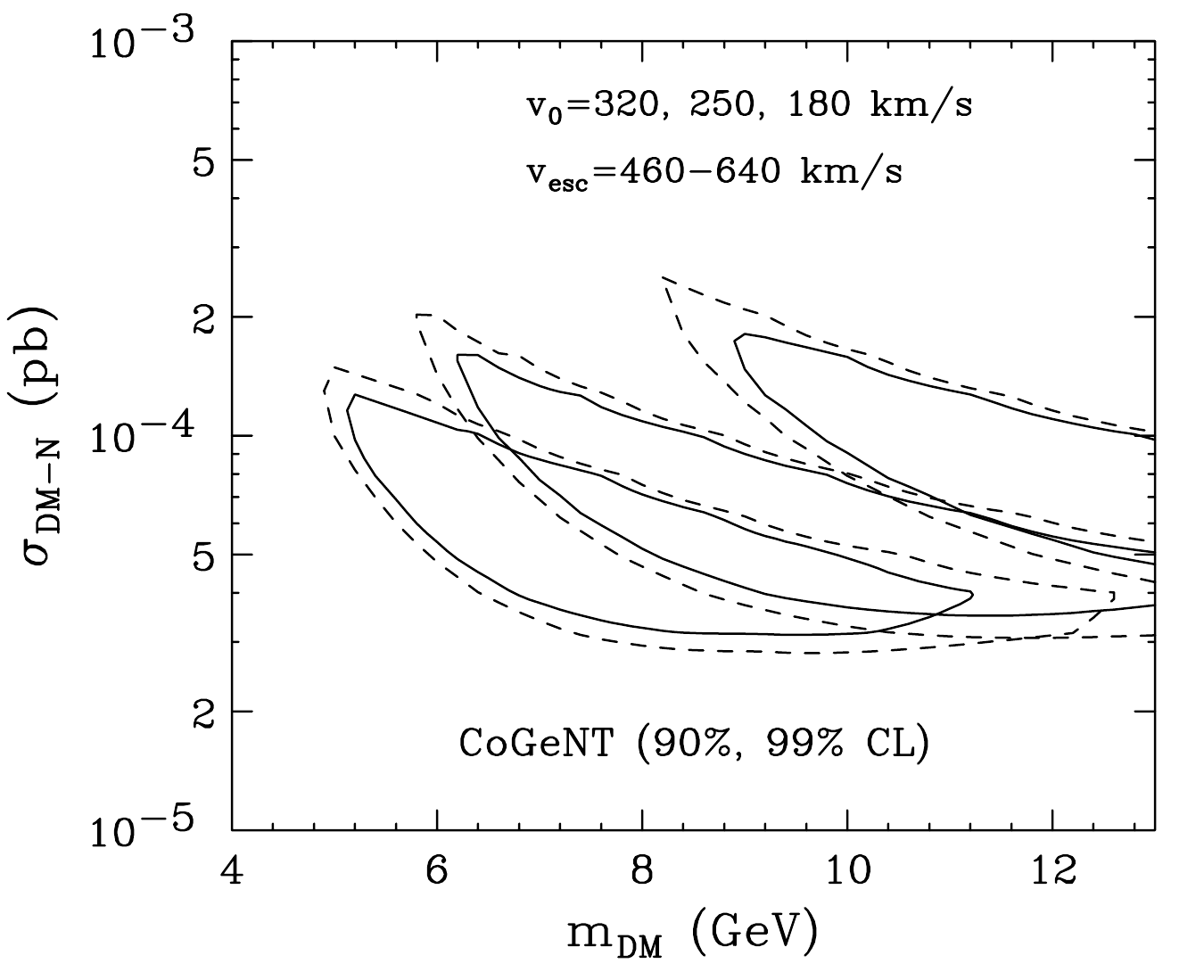}}
\caption{The range of dark matter parameter space that provides a good fit to the current CoGeNT data. The three overlapping regions correspond to $v_0 =320$, 250 and 180 km/s, from left-to-right. Each region has been marginalized over escape velocities between 460 and 640 km/s. The solid and dashed lines denote the 90\% and 99\% confidence level contours.}
\label{currentfit}
\end{figure}

Even with the relatively little exposure that went into this measurement (56 days, and 330 grams of target mass after cuts) CoGeNT has observed approximately $\sim$100 events between 0.4 and 1.0 keVee which cannot be accounted for by known backgrounds. In Fig.~\ref{currentfit}, we show the range of dark matter masses and cross sections that provide a good fit to the current CoGeNT excess, for various choices of the velocity distribution parameters. The three overlapping regions represent the range of mass and cross section that can fit the current data for values of $v_0=320, 250$ and 180 km/s, respectively (in each region, the escape velocity is marginalized over the range of 460 to 640 km/s).

\section{Future Projections For CoGeNT}
\label{future}

If the excess events reported by CoGeNT is the result of elastically scattering dark matter particles, then we should expect a degree of seasonal variation in the event rate. Due to the Earth's motion around the Sun, the rate of dark matter recoil events is predicted to vary throughout the year, peaking at or around June 2nd~\cite{modulation}. 

\begin{figure}[t]
\centering
{\includegraphics[angle=0.0,width=3.5in]{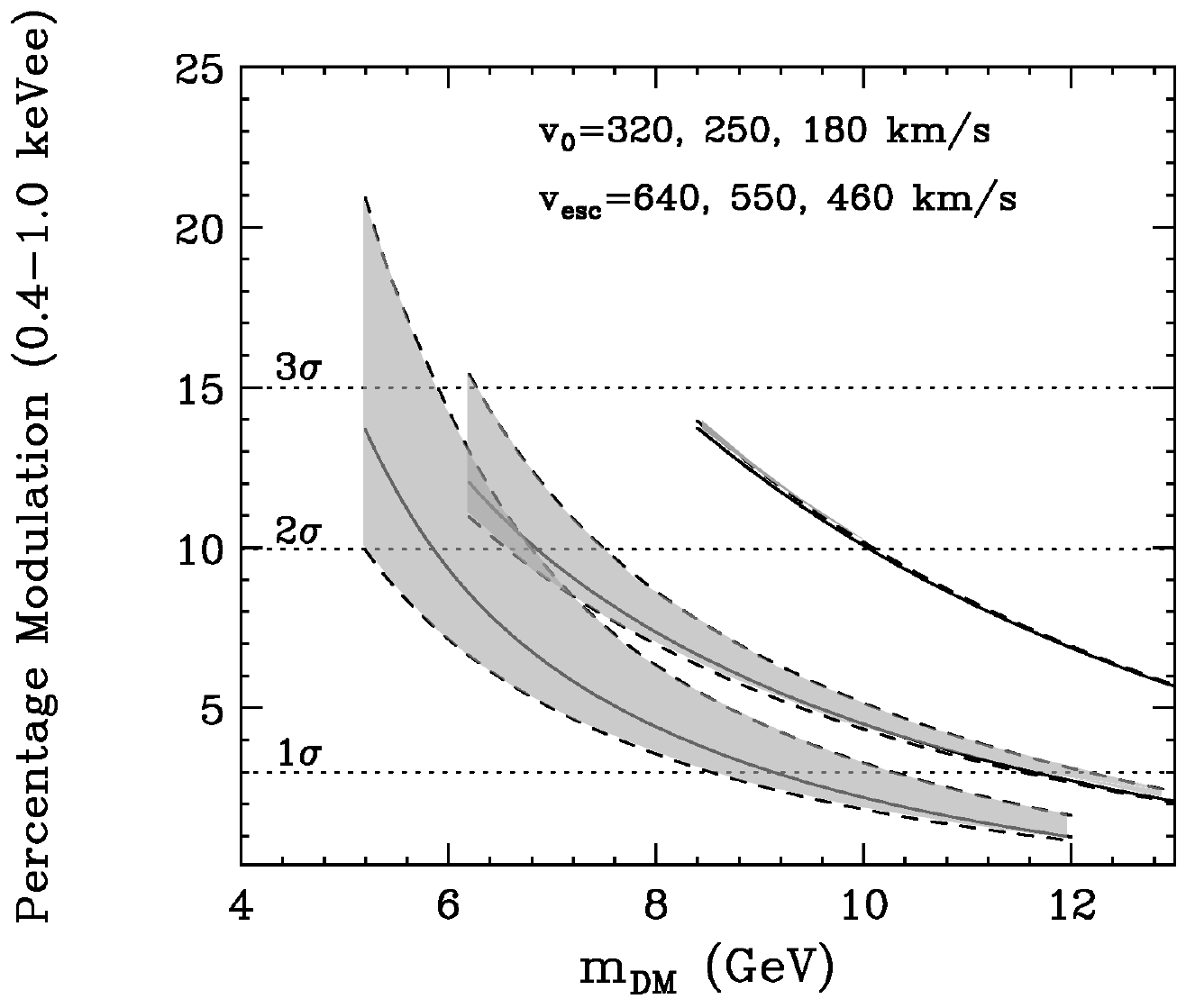}}
{\includegraphics[angle=0.0,width=3.5in]{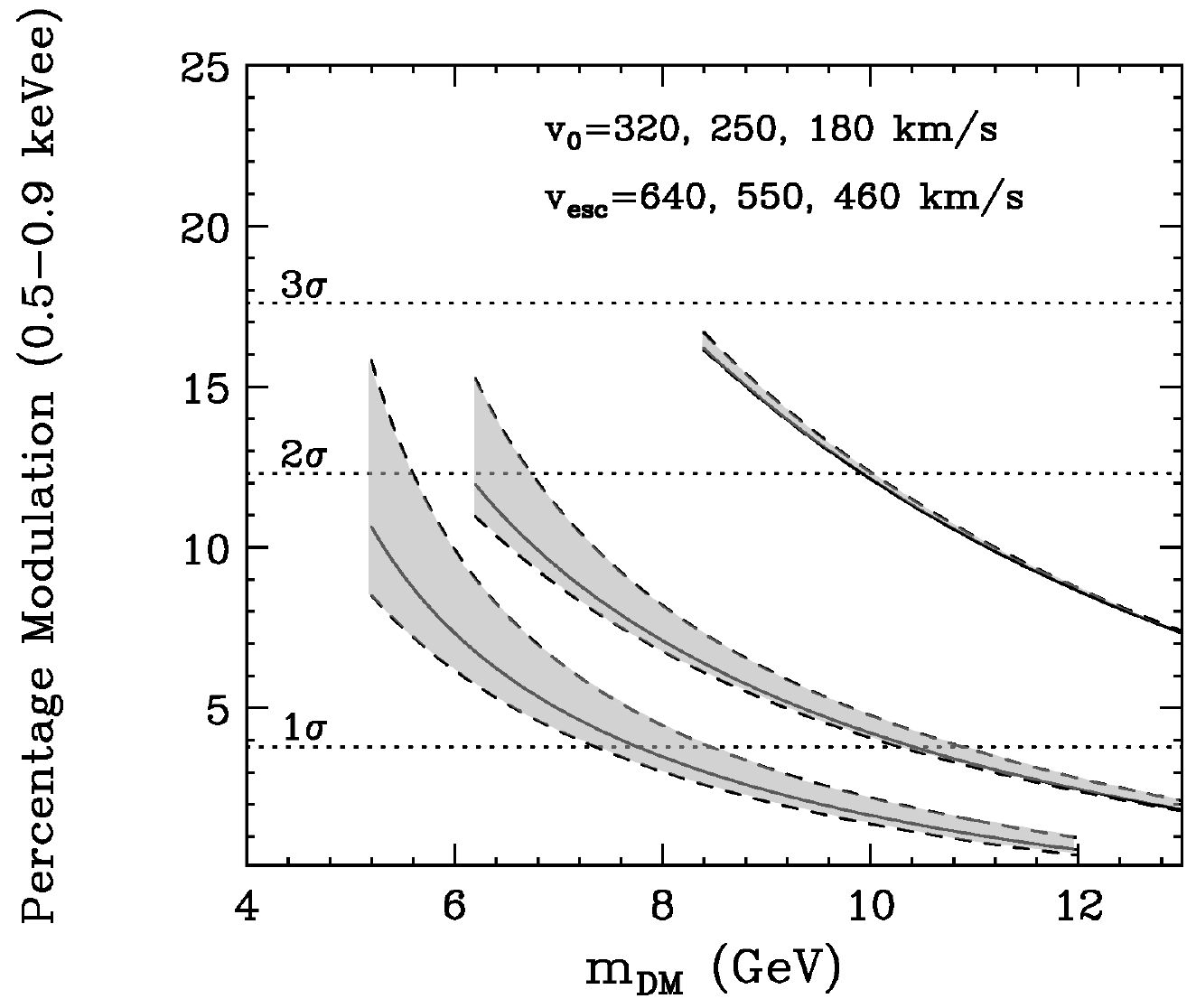}}
\caption{The predicted degree of annual modulation in the event rate of CoGeNT. The top (bottom) frame shows the modulation over the range of 0.4 to 1.0 keVee (0.5 to 0.9 keVee). The horizontal dotted lines denote the projected confidence levels at which CoGeNT will be able to identify a modulation with one full year of data. The three solid lines in each frame correspond to $v_0=320, 250,$ and 180 km/s, from left-to-right, each with $v_{\rm esc}=550$ km/s. The dashed lines above and below each solid line correspond to $v_{\rm esc}=640$ and 460 km/s, respectively.}
\label{mod}
\end{figure}

In Fig.~\ref{mod}, we show the predicted degree of modulation for dark matter within the parameter space region capable of generating the observed CoGeNT excess. The percentage modulation that is plotted is defined as $(R_{\rm summer}-R_{\rm winter})/(2 R_{\rm ave})$, where $R_{\rm summer}$ and $R_{\rm winter}$ denote the maxima and minima of the rate. The top frame of Fig.~\ref{mod} shows the modulation over the range of 0.4 to 1.0 keVee, while the bottom frame considers a more narrow (and more conservative) range of 0.5 to 0.9 keVee. In each frame, the horizontal dotted lines denote the projected confidence levels at which CoGeNT will be able to identify a modulation after one year.

\begin{figure}[t]
\centering
{\includegraphics[angle=0.0,width=3.5in]{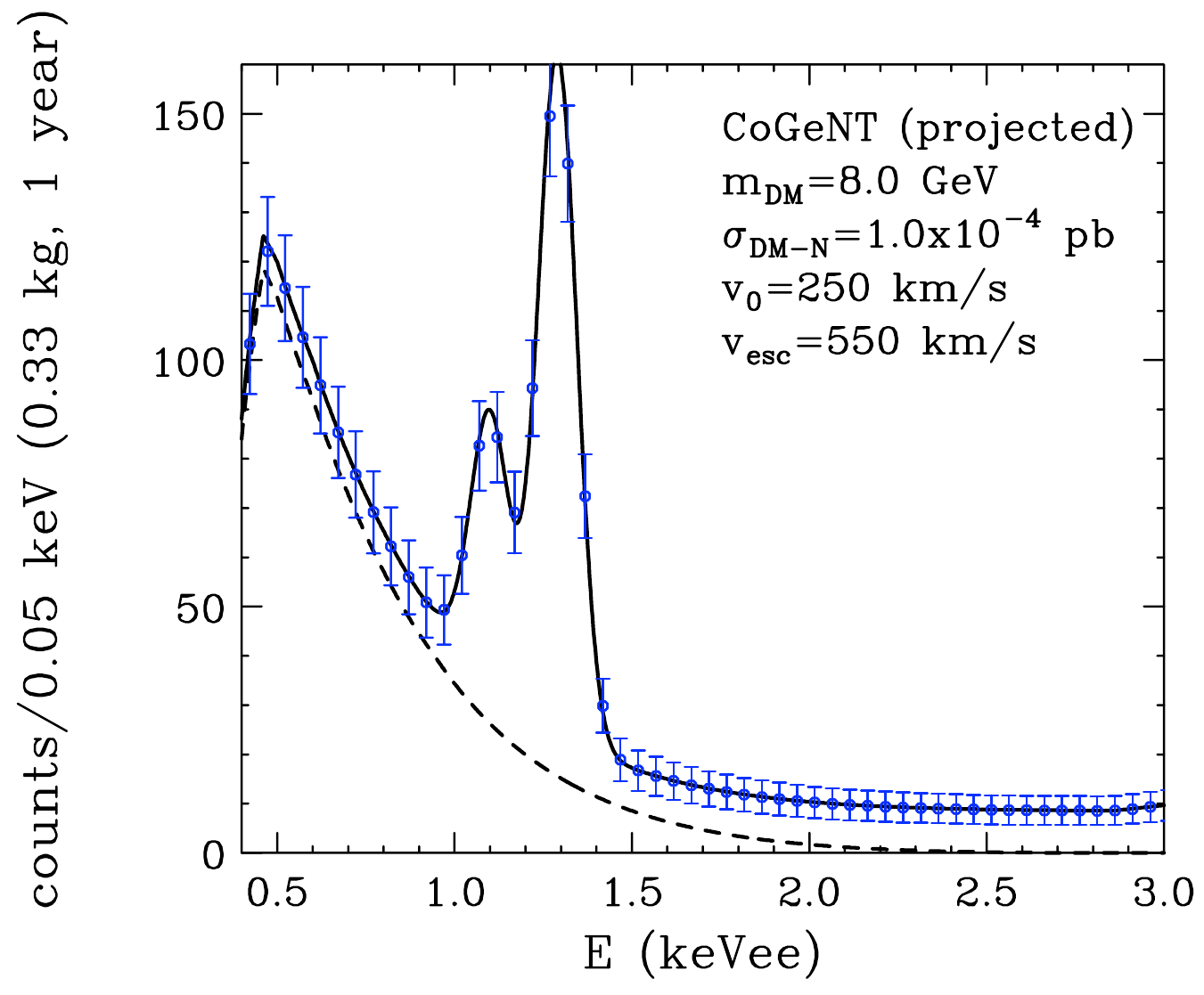}}\\
{\includegraphics[angle=0.0,width=3.5in]{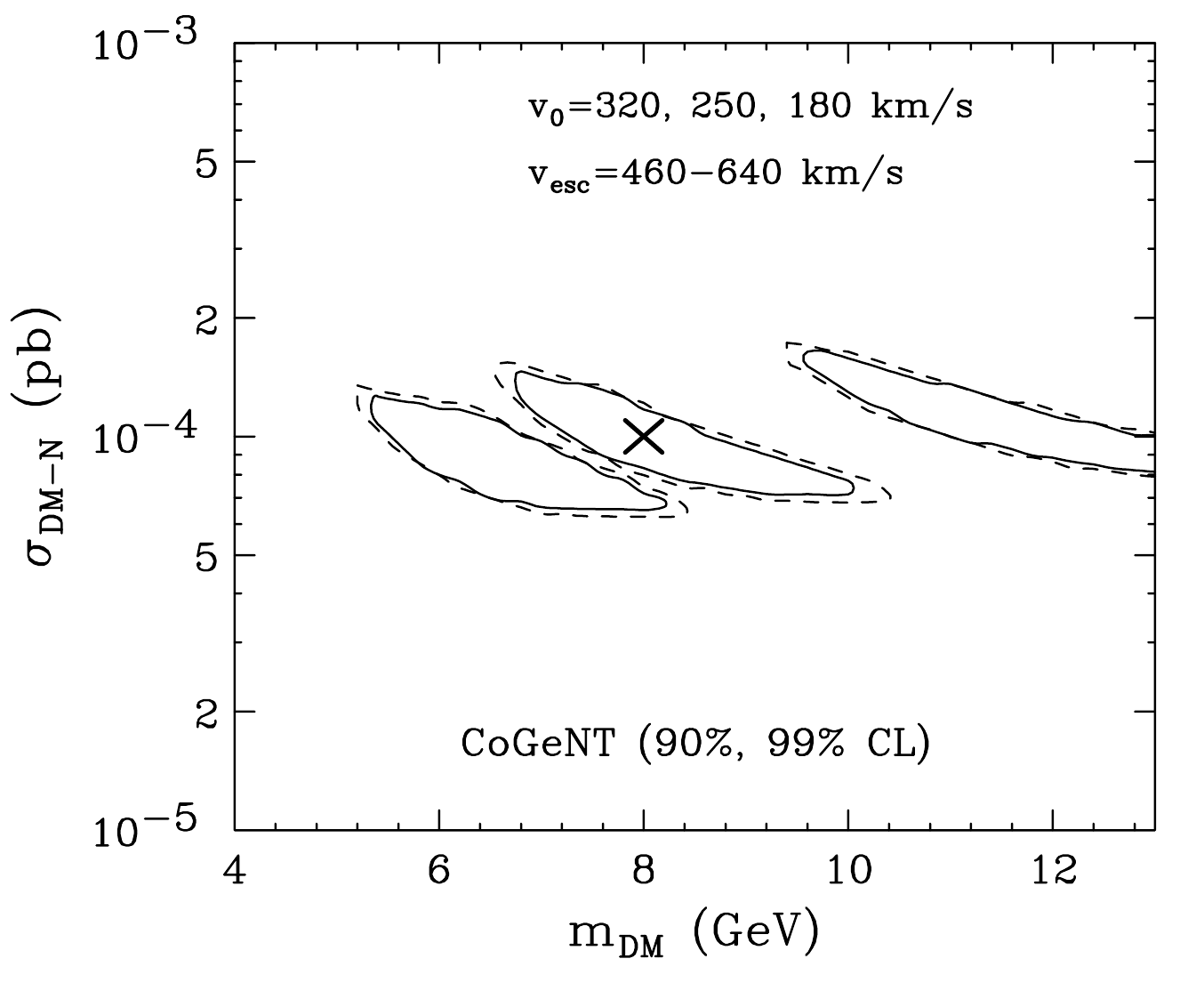}}
\caption{Top: The projected spectrum of events with one year of CoGeNT data, for an underlying dark matter model with a mass of 8 GeV, an elastic scattering cross section of $10^{-4}$ pb, $v_0=250$ km/s, and $v_{\rm esc}=550$ km/s (the ``true model''). Again, the dashed line denotes the spectrum of dark matter events alone, while the solid line is the dark matter spectrum plus backgrounds. Bottom: The range of dark matter parameter space that provides a good fit to the projected CoGeNT data. The three regions shown correspond to $v_0 =320$, 250 and 180 km/s, from left-to-right. Each region has been marginalized over escape velocities between 460 and 640 km/s. The solid and dashed lines denote the 90\% and 99\% confidence level contours. The $X$ denotes location of the true model.}
\label{oneyear}
\end{figure}

The additional data collected by CoGeNT over its first year will also considerably improve its ability to measure the spectrum of events, and place corresponding constraints on the underlying dark matter parameter space. In Fig.~\ref{oneyear}, we show the spectrum projected to be measured by CoGeNT after one year (top) and the corresponding regions of dark matter parameter space that provide a good fit to this spectrum (bottom). In calculating these regions, we considered only the spectrum between 0.4 and 1.0 keVee.

\begin{figure}[t]
\centering
{\includegraphics[angle=0.0,width=3.5in]{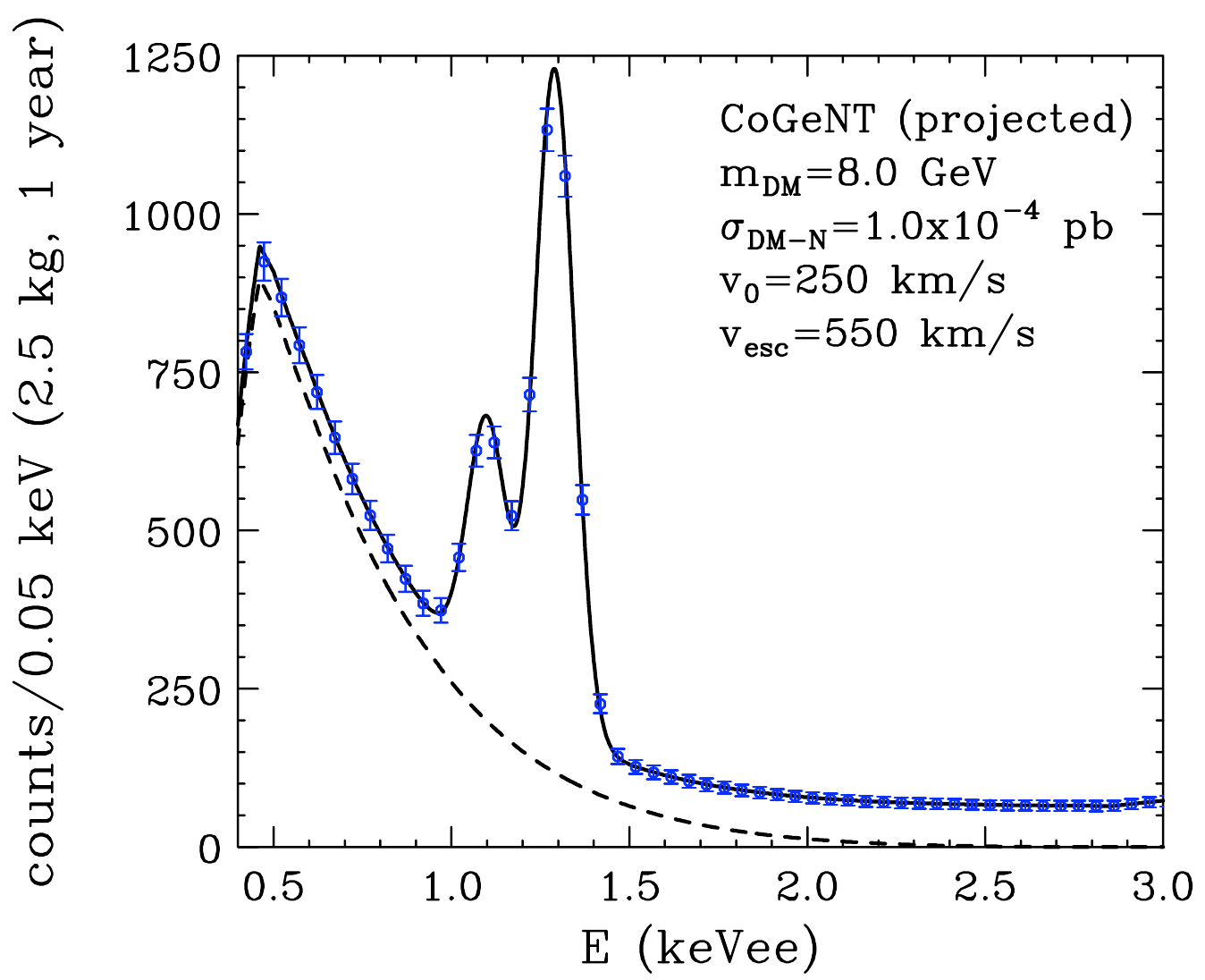}}\\
{\includegraphics[angle=0.0,width=3.5in]{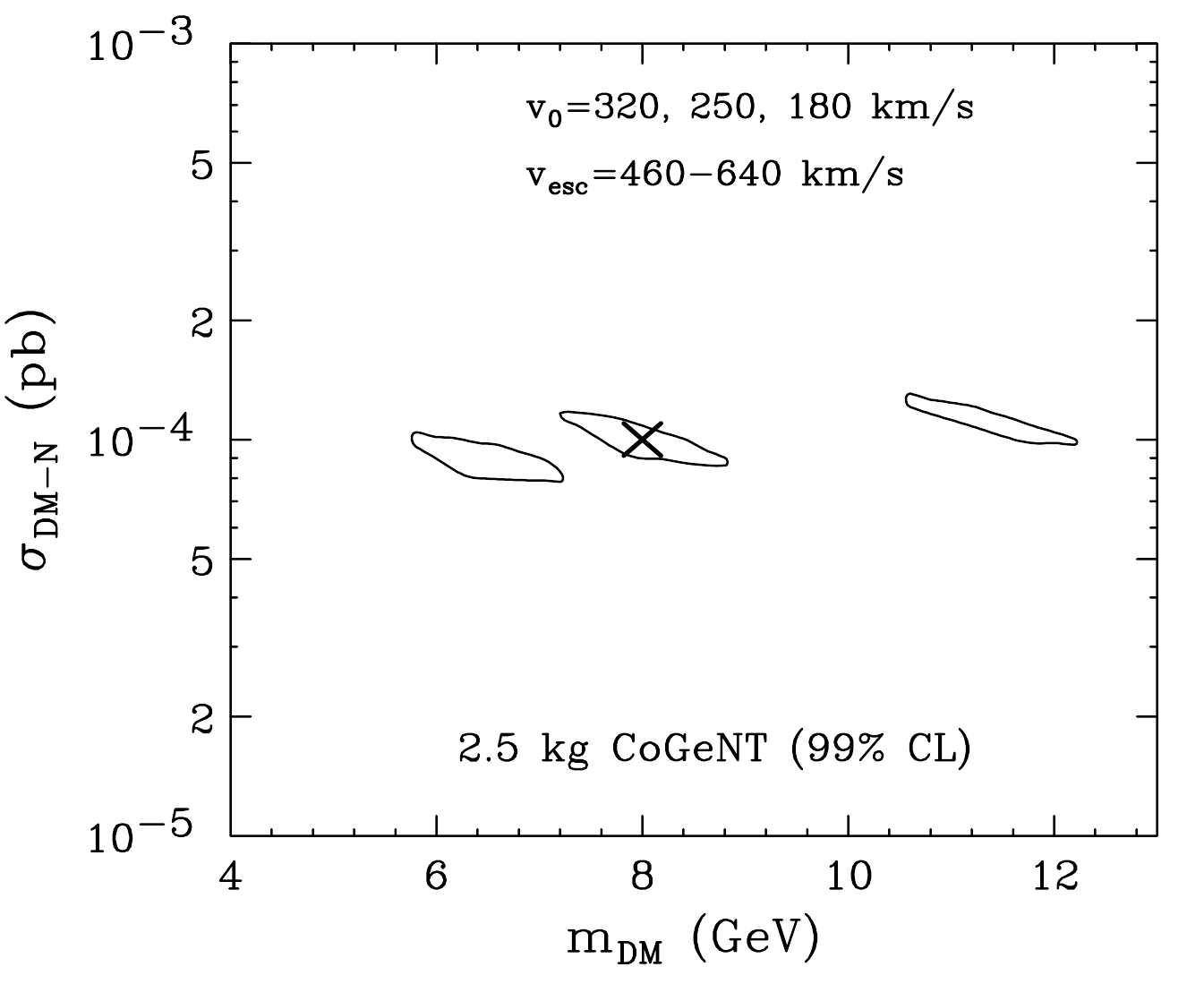}}
\caption{As shown in Fig.~\ref{oneyear}, but for an upgraded CoGeNT with a fiducial mass of 2.5 kg.}
\label{super}
\end{figure}

\begin{figure}[t]
\centering
{\includegraphics[angle=0.0,width=3.5in]{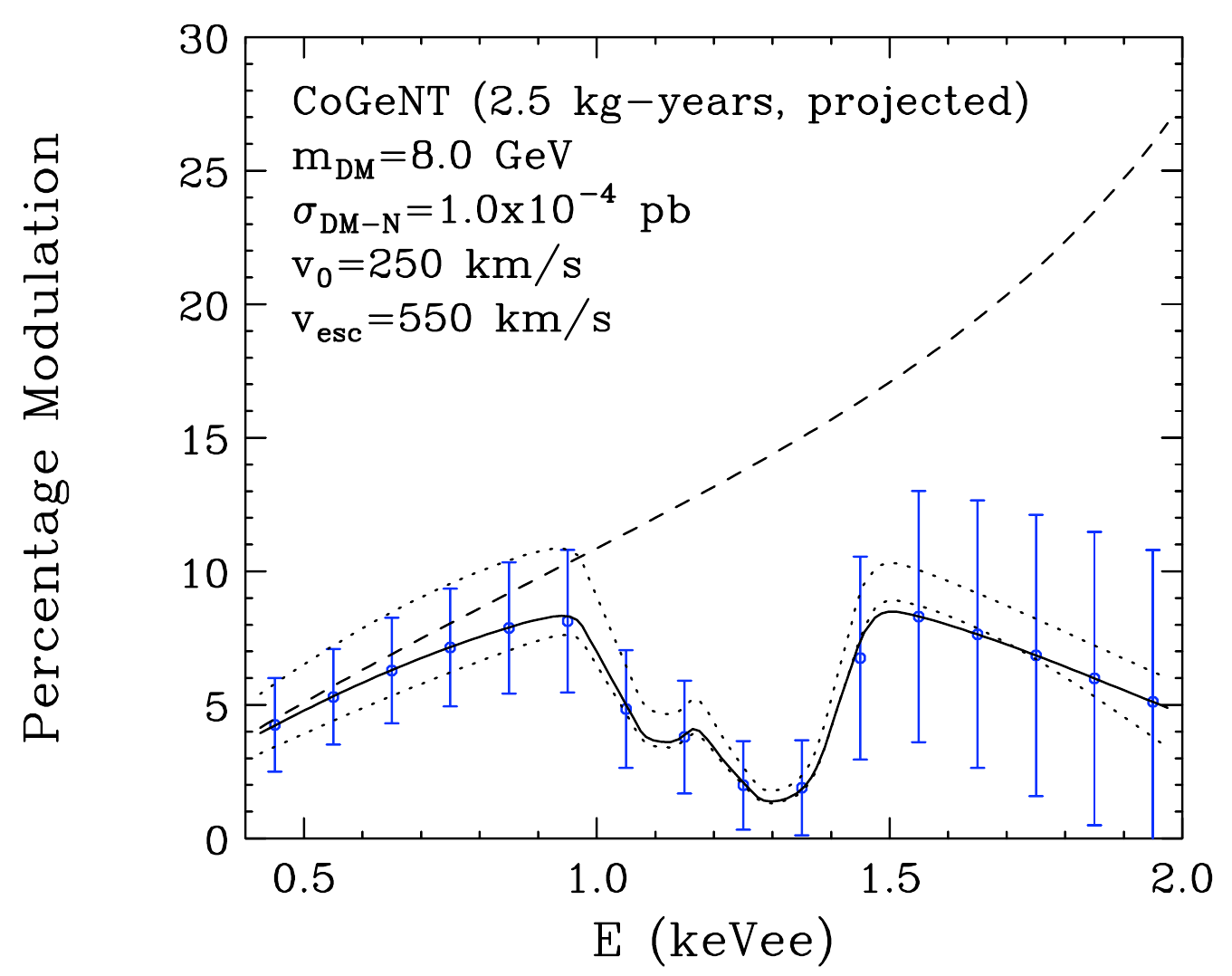}}
\caption{The measurement of the energy spectrum of the modulation, as projected for an upgraded CoGeNT with a 2.5 kg fidicial mass. The dashed line denotes the modulation for dark matter events alone, whereas the solid line represent the modulation in the total (signal plus background) rate. The upper (lower) dotted curve represents the prediction for parameters in the left (right) island-region in the bottom frame of Fig.~\ref{super}.}
\label{modspec}
\end{figure}

In Fig.~\ref{super}, we show similar results, but projected for the planned upgrade of CoGeNT, with a fiducial mass of 2.5 kg. In addition to improving upon the spectral constraints, such an upgrade will also make it possible to measure the energy spectrum of the modulation amplitude. We illustrate this in Fig.~\ref{modspec}. The solid curves shown in Fig.~\ref{modspec} denote the prediction for our central model ($m_{\rm DM}=$8 GeV, $\sigma_{\rm DM-N}=10^{-4}$ pb, $v_0=250$ km/s, $v_{\rm esc}=550$ km/s) including backgrounds, whereas the upper (lower) dotted curve represents the prediction for parameters in the left (right) island-region in the bottom frame of Fig.~\ref{super}. Note that this projection is somewhat conservative, as the 1.1 and 1.29 keV backgrounds are due to cosmogenic activation and will become steadily depleted with time.

\section{The Role of Other Experiments}
\label{others}

In this section, we will briefly discuss the ability of other direct detection experiments to study dark matter particles with the characteristics needed to generate the signals reported by CoGeNT and DAMA/LIBRA. Starting with DAMA/LIBRA, we note that although its comparatively enormous target mass (250 kg) has enabled it to make a very high significance measurement of the annual modulation amplitude, sizable uncertainties in the low energy behavior of its quenching factor limit the precision with which its data can be used to constrain the underlying dark matter parameter space. We also emphasize that, unlike CoGeNT, DAMA/LIBRA is capably only of measuring the amplitude of the modulation of their signal, and not the amplitude of the signal itself. The detection of a modulated rate by CoGeNT will provide the first combined measurement of modulation and signal amplitudes (the ratio of which is the percentage modulation shown in our figures), and will thus provide an important way of discriminating dark matter events from backgrounds. For an unknown background to modulate with a fractional amplitude that is similar to that predicted for dark matter would require an unfortunate and unlikely conspiracy of nature.

There are a number of other existing direct detection experiments that may be sensitive to 5-10 GeV dark matter particles. Very recently, the CDMS collaboration has used low-threshold data from their shallow Stanford site to place constraints on light dark matter particles~\cite{cdmslow}. At such low energies, however, the backgrounds observed by CDMS are not well understood, somewhat limiting their ability to probe the region implied by CoGeNT and DAMA/LIBRA. Dark matter experiments which use liquid Xenon as their target suffer from sizable uncertainties in the scintillation efficiency ($L_{\rm eff}$) and recoil energy scale ($Q_i$) at low recoil energies~\cite{uncertain}, making it difficult to interpret their results within the context of light dark matter.

Experiments with relatively light elements in their detectors, such as CRESST (CaWO$_4$) and COUPP (CF$_3$I) are potentially well suited for studying light dark matter. Over the past several months, the CRESST collaboration has described in talks an excess of events in their oxygen band~\cite{cresst} which could potentially arise from the elastic scattering of a dark matter particle in the CoGeNT/DAMA region~\cite{consistent}. We eagerly await further details pertaining to this observation. The COUPP collaboration has very recently begun operation of their 4 kg chamber at SNOLAB. A dark matter particle near the center of the region preferred by CoGeNT and DAMA/LIBRA is predicted to generate $\sim$1 event at COUPP per day (running with a recoil energy threshold of $\sim$7 keV). If their backgrounds are as low as anticipated, they could rapidly accumulate a significant excess of events.

\section{Summary and Outlook}
\label{summary}

The excess of low energy events reported by the CoGeNT collaboration (as well as the annual modulation reported by DAMA/LIBRA and the gamma ray emission observed from the Galactic Center) have generated a great deal of interest in 5-10 GeV dark matter particles. In this paper, we have studied the ability of current and future CoGeNT data to further elucidate this situation. Results from the first year of CoGeNT data are anticipated to be presented shortly. This data will not only enable the recoil energy spectrum of dark matter events to be measured with much greater precision, but will also likely be able to identify (or rule out) the presence of an annual modulation with a significance of 1-3$\sigma$. If observed, this would represent a major confirmation that CoGeNT's excess arises from dark matter. In the less immediate future, the planned upgrade for CoGeNT (with a fiducial mass of approximately 2.5 kg) will further improve upon these measurements, and will provide a measurement of the energy spectrum of the modulation amplitude.

\bigskip

As we were completing this study, the CDMS collaboration released a low threshold analysis of their Soudan data~\cite{newcdms}, which appears to conflict with the dark matter interpretation of the CoGeNT excess. Their data set, however, contains a large number of events classified as ``zero-charge'' background which overlap with the region of nuclear recoil candidates. A modest uncertainty or systematic error in the energy scale calibration at energies near their threshold could plausibly shift this exclusion contour above the region being considered here~\cite{jc}.

\section*{Acknowledgements} We would like to thank Juan Collar for helpful discussions. DH and CK are supported by the US Department of Energy, including grant DE-FG02-95ER40896, and by NASA grant NAG5-10842.


\begin{thebibliography}{9}



\bibitem{cogent}
C.~E.~Aalseth {\it et al.} [The CoGeNT Collaboration],
Phys.\ Rev.\ Lett.\, in press, 
arXiv:1002.4703 [astro-ph.CO]. 


\bibitem{zurek}
  A.~L.~Fitzpatrick, D.~Hooper, K.~M.~Zurek,
  Phys.\ Rev.\  {\bf D81}, 115005 (2010).
  [arXiv:1003.0014 [hep-ph]].


\bibitem{compare}
  S.~Chang, J.~Liu, A.~Pierce {\it et al.},
  [arXiv:1004.0697 [hep-ph]];
  S.~Andreas, C.~Arina, T.~Hambye {\it et al.},
  [arXiv:1003.2595 [hep-ph]];
  R.~Foot,
  [arXiv:1004.1424 [hep-ph]].
  






\bibitem{damanew}
  R.~Bernabei, P.~Belli, F.~Cappella {\it et al.},
  Eur.\ Phys.\ J.\  {\bf C67}, 39-49 (2010).
  [arXiv:1002.1028 [astro-ph.GA]].

 
  \bibitem{cdms}
 Z.~Ahmed {\it et al.} [The CDMS-II Collaboration],
  Science {\bf 327}, 1619-1621 (2010) [arXiv:0912.3592
[astro-ph.CO]]. 



\bibitem{xenon100}
  E.~Aprile {\it et al.} [XENON100 Collaboration],
  [arXiv:1005.0380 [astro-ph.CO]].



\bibitem{petriello}
  F.~Petriello and K.~M.~Zurek,
  JHEP {\bf 0809}, 047 (2008)
  [arXiv:0806.3989 [hep-ph]];
  C.~Savage, K.~Freese, P.~Gondolo and D.~Spolyar,
  JCAP {\bf 0909}, 036 (2009)
  [arXiv:0901.2713 [astro-ph]].
  C.~Savage, G.~Gelmini, P.~Gondolo and K.~Freese,
  JCAP {\bf 0904}, 010 (2009)
  [arXiv:0808.3607 [astro-ph]].
  








\bibitem{cdmslow}
CDMS Collaboration,
arXiv:1010.4290

\bibitem{constraints}
P.~Sorensen, talk given at IDM 2010, Montpellier;
 J.~Angle {\it et al.}  [XENON Collaboration],
  Phys.\ Rev.\ Lett.\  {\bf 100}, 021303 (2008)
  [arXiv:0706.0039 [astro-ph]];
  J.~Angle {\it et al.}  [XENON10 Collaboration],
  Phys.\ Rev.\  D {\bf 80}, 115005 (2009)
  [arXiv:0910.3698 [astro-ph.CO]];
  C.~Savage, G.~Gelmini, P.~Gondolo {\it et al.},
[arXiv:1006.0972v1 [astro-ph.CO]].


\bibitem{uncertain}
  J.~I.~Collar,
  arXiv:1010.5187 [astro-ph.IM];
  J.~I.~Collar, D.~N.~McKinsey,
  [arXiv:1005.0838 [astro-ph.CO]];
  [arXiv:1005.3723 [astro-ph.CO]];
  J.~I.~Collar,
  [arXiv:1006.2031 [astro-ph.CO]].

\bibitem{uncertain2}
  M.~Lisanti {\it et al.},
  [arXiv:1010.4300].


\bibitem{consistent}
  D.~Hooper, J.~I.~Collar, J.~Hall {\it et al.},
Phys.\ Rev.\  {\bf D}, in press, 
arXiv:1007.1005 [hep-ph].




\bibitem{Hooper:2010mq}
  D.~Hooper, L.~Goodenough,
  [arXiv:1010.2752 [hep-ph]].













\bibitem{ls}
  C.~Savage, K.~Freese, P.~Gondolo,
  Phys.\ Rev.\  {\bf D74}, 043531 (2006).
  [astro-ph/0607121]; 
See also J.~D.~Lewin and P.~F.~Smith,
  Astropart.\ Phys.\  {\bf 6}, 87-112 (1996).
  

\bibitem{velocity}
 M.~C.~Smith {\it et al.},
 Mon.\ Not.\ Roy.\ Astron.\ Soc.\  {\bf 379}, 755 (2007)
 [arXiv:astro-ph/0611671];
 R.~Schoenrich, J.~Binney and W.~Dehnen,
 arXiv:0912.3693 [astro-ph.GA];
 A.~M.~Green,
 JCAP {\bf 1010}, 034 (2010)
 [arXiv:1009.0916 [astro-ph.CO]];
 C.~McCabe,
 Phys.\ Rev.\  D {\bf 82}, 023530 (2010)
 [arXiv:1005.0579 [hep-ph]];
  P.~J.~Fox, J.~Liu and N.~Weiner,
  arXiv:1011.1915 [hep-ph].


\bibitem{Gondolo}
R. H. Helm, Phys. Rev. {\bf 104} 1466 (1956);
  G.~Duda, A.~Kemper and P.~Gondolo,
  JCAP {\bf 0704}, 012 (2007)
  [arXiv:hep-ph/0608035].

\bibitem{Fricke}
  G.~Fricke, C.~Bernhardt, K.~Heilig, L.~A.~Schaller,
L.~Schellenberg, E.~B.~Shera and C.~W.~de Jager,
  Atom.\ Data Nucl.\ Data Tabl.\  {\bf 60}, 177 (1995).






\bibitem{Ge}
For a summary of germanium quenching factor measurements, see
  S.~T.~Lin {\it et al.} [TEXONO Collaboration],
  Phys.\ Rev.\  {\bf D79}, 061101 (2009).
  [arXiv:0712.1645 [hep-ex]];
  Y.~Messous,
  Astropart.\ Phys.\  {\bf 3}, 361-366 (1995).
  
\bibitem{Ge2}
  P.~S.~Barbeau, J.~I.~Collar and O.~Tench,
  JCAP {\bf 09}, 009 (2007); P.~S.~Barbeau, J.~I.~Collar and 
  P.~M.~Whaley, Nucl. \ Instr. \ Meth.\ {\bf A574}, 385-391 (2007);
  P.~S.~Barbeau, PhD Thesis, University of Chicago (2009).
  
  


    
\bibitem{modulation}
  A.~K.~Drukier, K.~Freese, D.~N.~Spergel,
  Phys.\ Rev.\  {\bf D33}, 3495-3508 (1986).
  


\bibitem{cresst}
See talks by W.~Seidel, WONDER 2010 Workshop, Laboratory Nazionali del
Gran Sasso, Italy, March 22-23, 2010, and IDM 2010 Workshop, Montpellier, France, July 26-30, 2010, and MPIK seminar by T.~Schwetz, June 21, 2010.



\bibitem{newcdms}
  Z.~Ahmed, {\it et al.} [CDMS Collaboration],
  [arXiv:1011.2482 [astro-ph.CO]].


\bibitem{jc}
J.~Collar, private communication.

\end{thebibliography}
\end{document}